\documentclass[12pt]{article}

\usepackage[normalem]{ulem}
\usepackage{lmodern}
\usepackage{subfigure}
\usepackage{caption}
\usepackage{amsthm}
\usepackage{amsmath}
\usepackage{animate}
\usepackage{graphics}
\usepackage[a4paper, total={6in, 8in}]{geometry}

\usepackage{enumitem}
\usepackage{multicol}
\usepackage{graphicx}
\usepackage{epstopdf}
\usepackage{tabularx}
\usepackage[linesnumbered,titlenumbered,ruled,]{algorithm2e}

\usepackage{listings}
\usepackage{lmodern}
\usepackage[usenames,dvipsnames]{xcolor}

\usepackage{algorithmicx}
\usepackage{algpseudocode}
\usepackage{xcolor}

\usepackage{url}

\usepackage[toc,page]{appendix}

\algrenewcommand\alglinenumber[1]{\tiny #1:}
\usepackage{mathtools}
\usepackage{amsmath}%
\usepackage{amsfonts}%
\usepackage{amssymb}%
\usepackage{graphicx}
\usepackage{color}
\usepackage{cases}

\SetAlCapNameFnt{\scriptsize}
\SetAlCapFnt{\scriptsize}
\SetAlgorithmName{Algorithm}{Algo}{List of Algorithms}

\newlist{myenumi}{description}{10}
\setlist[myenumi]{labelindent=\parindent, leftmargin=*, label=(\roman*), align=left}
\setlist[myenumi]{leftmargin=0pt}

\newtheorem{innercustomgeneric}{\customgenericname}
\providecommand{\customgenericname}{}
\newcommand{\newcustomtheorem}[2]{%
  \newenvironment{#1}[1]
  {%
   \renewcommand\customgenericname{#2}%
   \renewcommand\theinnercustomgeneric{##1}%
   \innercustomgeneric
  }
  {\endinnercustomgeneric}
}

\newcustomtheorem{conj}{Conjecture}
\newcustomtheorem{customlemma}{Lemma}

\setlist[itemize]{noitemsep, topsep=0pt}

    \newcommand{\iom}{$IO\mathcal{M}$\xspace}
    \newcommand{\fom}{$FO\mathcal{M}$\xspace}
    \newcommand{\foj}{$FO\mathcal{K}_c$\xspace}
    \newcommand{\ioj}{$IO\mathcal{K}_c$\xspace}
    \newcommand{\D}{\Delta }
\newtheorem{theorem}{Theorem}
\newtheorem{remark}[theorem]{Remark}

\newtheorem{proposition}[theorem]{Proposition}

    \definecolor{darkgreen}{rgb}{0.1, 0.5, 0.1}
\begin{document}

\title{Mandelbrot set as a particular Julia set of Fractional Order, echipotential lines and external rays of Mandelbrot and Julia sets of Fractional Order
}


\author{Marius-F. Danca{\footnote{Corresponding author}}\\
STAR-UBB Institute,\\
Babes-Bolyai University,\\
400084, Cluj-Napoca, Romania\\
Email: m.f.danca@gmail.com\\
}

\maketitle
\begin{abstract}
In this paper it is shown analytically and computationally that the Mandelbrot set of integer order are particular cases of Julia sets of Caputo's like fractional order. Also the differences between the fractional-order Mandelbrot set and Julia sets from their integer-order variants are revealed. Equipotential lines and external rays of Mandelbrot set and Julia sets of fractional order are determined.
\end{abstract}


Fractional-order Mandelbrot and Julia sets defined in the sense of $q$ -th Caputo-like discrete fractional differences

\section{Introduction}

The fractional-order (FO) Mandelbrot and Julia sets in the sense of $q$-th Caputo-like discrete fractional differences, for $q \in (0, 1)$, generated by the quadratic complex (Mandelbrot) map $f_c(z)=z^2+c$, with $z$ and $c$ complex and starting from the initial value $z_0=0$ (the critical point), are introduced in \cite{danca1} (see also \cite{plu15,oxi,oxi2,oxi3}). The algorithms for generating FO sets, base on the known Mandelbrot set and Julia sets of Integer Order (IO) which, after they were discovered, still represent a huge source of inspiration for computer graphics programmers but also for mathematicians. The first who drawn the Mandelbrot set of IO are Robert W. Brooks and Peter Matelski in 1978 \cite{m1} before the American-French-Polish mathematician Benoit B. Mandelbrot made it famous and given him the importance and the place in the chaos theory \cite{mana1}. Moreover, the Mandelbrot set, an invariant universal set, is considered to play a similar role as $\pi$ and $e$ have in mathematics. History of the origin of the Mandelbrot can be found at \cite{wiki}. Before the born of the Mandelbrot set, the study of the dynamics of complex maps was initiated by P. Fatou and G. Julia in the early of twentieth century \cite{juli,fat}.
 For fractal structures see e.g. \cite{bibzece,bibunspe,sci,mandelus}, while for details and background on Mandelbrot set and Julia sets see \cite{mand2,dou,bibcinspe,bibzece,devu}. Mandelbrot set can be considered as a book with an infinity of pages, each page being a Julia set.

 While to generate the Mandelbrot set, $c$ is considered variable within a lattice in the parametric plane $\mathbb{C}$, the Julia sets are obtained with fixed $c$, the origin of iterations of $f_c(z)$ being variable in the considered lattice.

The infinite beauty of these fractal sets, generated by the quadratic map, does not represent the subject of this paper, interested readers being directed to, e.g., \cite{bibzece,sci,mandelus}.

It was found that the fractional calculus represents more accurate the natural behavior in areas bioengineering, recurrent neural network, of engineering, image encryption, electronics, viscoelasticity, robotics, control theory and so on (see e.g. \cite{plu1,plu2,plu4,plu5,mumu,mumu2,mumu3}).
First definitions of a fractional difference operator are proposed in 1974 \cite{4}.
Aspects related to Caputo fractional sums and differences can be found in \cite{6,mich}, while Initial Value Problems (IVPs) in fractional differences are studied in \cite{bibnoua}. The stability of fractional differences is snalyzed in \cite{9,10}, and weakly fractional difference equations, symmetry-breaking of fractional maps can be found in \cite{x1,x2,bib1}. For nonexistence of periodic solutions see \cite{micy}.

In this paper the fractional Mandelbrot and Julia sets are introduced and some important properties are analytically and computationally presented. The discrete fractional calculus is used as a natural extension of difference calculus and Mandelbrot's idea of creating fractals from iterative of complex mappings to study iterations of complex fractional difference equations is extended.

Notations utilized in this paper:
\begin{itemize}
\item \iom: Mandelbrot of IO;
\item $IO\mathcal{K}_c$: Filled Julia set of IO;
\item $IO\mathcal{J}_c$: Julia set of IO;
\item \fom: Mandelbrot set of FO;
\item \foj: Filled Julia set of FO;

\end{itemize}

\section{Mandelbrot set and Julia sets of FO}
Next a briefly recall of some elementary notions about Mandelbrot set and Julia sets of IO required by the FO counterparts is presented (see \cite{bibzece,sci} or \cite{dancu} for more information).

The iteration of $f_c$ with $z(0)=0$
\begin{equation}\label{zero}
z_{n}=f_c(z_{n-1})=z_{n-1}^2+c, z_0=0, n\geq1,
\end{equation}
generates the sequence
\begin{equation}\label{unu}
z_0=0,z_1=f_c(0)=c,z_2=f_c^2(0)=c^2+c,z_3=f_c^3(0)=(c^2+c)^2+c,...,
\end{equation}
which will be used to generate Mandelbrot sets, while for $z(0)\neq0$, the sequence of iterates becomes
\begin{equation}\label{unus}
z_0,z_0^2+c,(z_0^2+c)^2+c,((z_0^2+c)^2+c)^2+c,...
\end{equation}
used to generate Julia sets.

The \emph{Mandelbrot set} of IO, \iom,  is the set of complex values $c$ for which the absolute value of $z_n$ remains bounded and does not tend to infinite, for all $n\geq1$, $|z(n)|<M$, usually $r$ taken as $r=M$ \cite{sci}, but could be taken even of order of thousands.

To define, for a fixed $c$, the \emph{Julia sets} of IO, $IO\mathcal{J}_c$, let consider the attraction basin of $\infty$,  $A_c(\infty)$, the set of points $z_0$ which tend to $\infty$ through the iteration \eqref{zero}
\[
A_c(\infty)=\{z_0\in\mathbb{C}: f_c^k(z_0)\rightarrow \infty, \text{~as~} k\rightarrow \infty\}.
\]
The boundary of $A_c(\infty)$, which depends on $c$, represents the \emph{Julia set }of IO, $IO\mathcal{J}_c$  \cite{sci}

\[
IO\mathcal{J}_c=\partial A_c(\infty)
\]

Another notion related to Julia sets of IO considered in this paper, is the \emph{filled Julia set} of IO, $IO\mathcal{K}_c$, which for a fixed $c$, is the set of all points $z_0\in \mathbb{C}$ for which the orbit \eqref{unu} remains bounded
\[
IO\mathcal{K}_c=\{z_0\in \mathbb{C}: f_c^k(z_0) \text{~~remains bounded for all~} k\}=\mathbb{C}\backslash A_c(\infty).
\]

The $IO\mathcal{J}_c$ set is contained in the $IO\mathcal{K}_c$ set and is the boundary of the $IO\mathcal{K}_c$ set \cite{sci}
\[
\partial IO\mathcal{K}_c=IO\mathcal{J}_c=\partial A_c(\infty).
\]

In this paper for computer graphic reasons, without loss of generality, one consider the analysis and computations of the \ioj sets.

To draw the \iom and $IO\mathcal{K}_c$ the so called \emph{escape-time algorithm } (\emph{direct algorithm}) is used \cite{dancu}. For Mandelbrot set, $z_0 = 0$, and $c$ is varied within a finite complex parametric domain (lattice), while for Julia sets, $z_0$ is varied within a finite lattice for fixed $c$. If after a finite number of iterations $N$ of $f_c$ the modulus $|z_n|$, $n=1,2,...,N$, remains bounded ($|z_N|\leq2$), then $c$ in the case of \iom set, or $z_0$ in the case of $IO\mathcal{K}_c$ sets, belongs to the \iom or $IO\mathcal{K}_c$, respectively. Otherwise, $c$ or $z_0$ does not belongs to \iom or $IO\mathcal{K}_c$, respectively.

In the graphical representations of this paper, the complex plane of Mandelbrot sets is the plane $(c_x,c_y)$ (the parameter space), where $c_x$ and $c_y$ are the coordinates of $c$, while the Julia sets are drawn in the complex plane $(x,y)$ of the initial point $z(0)$ of coordinates $x$ and $y$.
The sets \iom and  $IO\mathcal{K}_c$ are usually plotted as color, most often black, while outside points of these sets can be plotted colored using smooth coloring schemes or black-white \cite{bibzece}, \cite{ur} (see Fig. \ref{fig1} (b)  for the \iom and Fig. \ref{fig1}  (c) and (d) for \ioj sets generated for $c$ considered as points $A$ and $B$ in the \iom set).

Some of the most important properties of the \iom set and $IO\mathcal{K}_c$ sets analyzed in this paper are :

\begin{itemize}
\item [P1.] B. B. Mandelbrot found empirically that some isolated islands out of the body of the \iom set were actually connected to the mainland by very thin filaments \cite{mil} (Fig. \ref{fig1} (b). Now, it is conjectured that the Mandelbrot set is locally connected \cite{dou,wiki};
\item [P2.] \ioj sets are connected if the underlying $c$ belong to the interior of the \iom, i.e. \iom set is the set of all parameters $c$ for which \ioj is a connected set;
\item [P3.] For $c$ chosen at the boundary of the Mandelbrot set the related \ioj set is a ``dendrite'', while for $c$ situated outside of \iom the corresponding \ioj is a Cantor sets, or ``dust''-like, composed of infinitely many disjoint points \cite{bibunspe} (see Fig. \ref{fig1} (c) and (d) where the \ioj sets corresponding to points $A$ and $B$ are presented).
\end{itemize}

\section{Mandelbrot and Julia maps of FO}\label{sect}

Let the time scale $N_a=\{a,a+1,a+2,...\}$. The $q$-th Caputo-like discrete fractional difference of a function $u:\mathbb{N}_a\rightarrow \mathbb{R}$, for $q>0$ and $q\not \in \mathbb{N}$, is defined as is defined \cite{bibcinci} as
\begin{equation*}\label{capa1}
\Delta_a^q u(t)=\Delta_a^{-(n-q)}\Delta^n u(t)=\frac{1}{\Gamma(n-q)}\sum_{s=a}^{t-(n-q)}(t-s-1)^{(n-q-1)}\Delta^nu(s),
\end{equation*}
where $t\in \mathbb{N}_{a+n-q}$ and $n=[q]+1$.

$\Delta^n$ is the $n$-th order forward difference operator,
\[
\Delta^n u(s)=\sum_{k=0}^{n}\binom {n}{k}(-1)^{n-k}u(s+k),
\]
while $\Delta_a^{-q}$ represents the fractional sum of order $q$ of $u$, namely,
\begin{equation*}\label{suma}
\Delta_a^{-q}u(t)=\frac{1}{\Gamma(q)}\sum_{s=a}^{t-q}(t-s-1)^{(q-1)}u(s),~t\in \mathbb{N}_{a+q}.
\end{equation*}
The falling factorial $t^{(q)}$, is defined as follows:
\[
t^{(q)}=\frac{\Gamma(t+1)}{\Gamma(t-q+1)}.
\]

Note that the fractional operator  $\D_a^{-q}$ maps functions on $\mathbb{N}_a$ to functions
on $\mathbb{N}_{a+q}$ (for time scales see, e.g., \cite{time}).

For $q\in(0,1)$, when $\D u(s)=u(s+1)-u(s)$, $n=1$, and starting point $a=0$, case considered in this paper, $q$-th Caputo's difference, $\D ^q$, becomes
\begin{equation*}\label{capa}
\D^q u(t)=\frac{1}{\Gamma(1-q)}\sum_{s=a}^{t-(1-q)}(t-s-1)^{(-q)}\Delta u(s).
\end{equation*}
Then, the real FO autonomous Initial Value Problem (IVP) in the sense of Caputo
\begin{equation*}\label{trei}
\D^q u(t)=f(u(t+q-1)),~~ t\in \mathbb{N}_{1-q},~~u(0)=u_0,
\end{equation*}
with $f$ a continuous real valued map and $q\in(0,1)$, has the numerical solution
\begin{equation*}\label{inte}
u(t)=u_0+\frac{1}{\Gamma(q)}\sum_{s=1-q}^{t-q}(t-s-1)^{(q-1)}f(u(s+q-1)),
\end{equation*}
with the commonly used form in numerical applications
\begin{equation}\label{eq3}
u(n)=u(0)+\frac{1}{\Gamma(q)}\sum_{i=1}^n\frac{\Gamma(n-i+q)}{\Gamma(n-i+1)}f(u(i-1)),~{n\in \mathbb{N}^*},
\end{equation}
Consider next the complex variant of the IVP of FO

\begin{equation*}\label{primus}
\Delta^q z(t)=f_c(z(t+q-1),~~t\in \mathbb{N}_{1-q}, ~~z(0)=z_0,
\end{equation*}
with $q\in(0,1)$, $z=x+\i y \in \mathbb{C}$,  scaled $c$ within a parametric complex domain, and $z_0\in \mathbb{C}$. Then, the numerical integral \eqref{eq3} becomes \cite{dancu}

\begin{equation}\label{ecuss}
 z(n)=z(0)+\frac{1}{\Gamma(q)}\sum_{i=1}^n\frac{\Gamma(n-i+q)}{\Gamma(n-i+1)}f_c(z(i-1)),~ {n\in \mathbb{N}^*},
\end{equation}

\noindent which represents the mathematical description of the complex logistic (Mandelbrot) map of FO used to generate the \fom set (with $z(0)=0$) or \foj sets (with $z(0)$ variable).

\begin{remark}
To note that whole for the \iom (\ioj) set one iterates $f_c$, usually few dozens iterations, $N$, to generate the \fom (\foj) set, every iteration of  \eqref{ecuss} necessary to obtain the set, the expression of $z(n)$ requires the calculation of the sum in the right hand side of \eqref{ecuss} for each $n=1,2,...,N$. Only small values of $N$, of order of few tens (e.g. $N=20$ as for the \iom set) does not provide good accuracy in the calculation of $z(n)$. Therefore, a compromise between a relatively higher $N$ (time consuming), e.g. $N=70-100$, and computing time is desirable.
\end{remark}

To obtain the \fom set, one iterates \eqref{ecuss}, for $c$ scanning a complex domain (usually a rectangular lattice \cite{dancu}). The set of points $c$ for which the sequence of modules $|z(n)|$ remains bounded after a finite number of iterations $N$, forms the \fom set.

To obtain the \foj sets one fix $c$ (see P2 and P3) and one iterates \eqref{ecuss} with $z(0)$ variable within a complex domain. Like for the \fom set, if after $N$ iterations $|z(n)|$ remains bounded, $z(0)$ belongs to the underlying \foj set. In Fig. \ref{fig2} (b) is drawn the \fom for $q=0.5$, while in Figs. \ref{fig2} (c),(d) and (e) several \foj sets are drawn for $q=0.5$ and $c=-0.0781+0.6694i$, $c=-1.0516+0.0913i$, $c=-0.5$, respectively. Details on FO algorithms can be found in \cite{dancu}, while a matlab code for \fom can be found at \cite{cod}.

\section{Properties of the \fom set}

Several properties of the \fom map are analyzed in \cite{dancu}, especially for real $c$. In this paper some properties are analytically proved and verified with aided of the scientific computation.

Contrary to expectations, the \iom set is not a particular case of the \fom for $q=1$, but only for $q\downarrow 0$ as proved bellow

\begin{proposition}\label{propi1}The \iom set is the \fom set for $q\downarrow 0$.
\begin{proof}
Consider the limit of $z(n)$ for $q\downarrow 0$, with $z(0)=0$ in \eqref{ecuss} rearranged as follows
\begin{equation*}
\begin{split}
\lim_{q\downarrow 0}z(n)=&\lim_{q\downarrow 0}\bigg(\frac{1}{\Gamma(q)}\sum_{i=1}^{n}\frac{\Gamma(n-i+q)}{\Gamma(n-i+1)}f_c(z(i-1))\bigg)\\
&=\lim_{q\downarrow 0}\frac{1}{\Gamma(q)}\sum_{i=1}^{n-1}\frac{\Gamma(n-i+q)}{\Gamma(n-i+1)}f_c(z(i-1))+\lim_{q\downarrow 0}\frac{1}{\Gamma(q)}\frac{\Gamma(q)}{\Gamma(1)}f_c(z(n-1)).
\end{split}
\end{equation*}
Because
\[
\lim_{q\downarrow 0}\frac{1}{\Gamma(q)}=0,
\]
and for a finite times of iterations, $\sum_{i=1}^{n-1}\frac{\Gamma(n-i+q)}{\Gamma(n-i+1)}$ is bounded, it follows that
\[
\lim_{q\downarrow 0}\frac{1}{\Gamma(q)}\sum_{i=1}^{n-1}\frac{\Gamma(n-i+q)}{\Gamma(n-i+1)}=0,
\]
and, therefore,
\[
\lim_{q\downarrow 0}z(n)=f_c(z(n-1)),
\]
or, in simplified form
\[
z(n)=f_c(z(n-1)),
\]
i.e. the \iom map.
\end{proof}
\end{proposition}

 Because at $q=0$, $\Gamma(q)$ has a simple pole singularity, in simulations $q$ cannot be set $0$ and, therefore, it is considered numerically, e.g., $q=10^{-m}$ with $m$ some positive integer. In Fig. \ref{fig2} (a) the \fom set is presented for $m=6$.

 An animation showing the metamorphosis of the \fom sets for $q$ varying from $q=1$ to $q=0$ is presented as a supplementary video.

For $c\neq0$, Proposition \ref{propi1} no longer takes place for the \foj map, because in this case one obtains:
\[
\lim_{q\downarrow 0}z(n)=z(0)+f_c(z(n-1)),
\]
which by iteration, generates the sequence
\begin{equation}\label{jj0}
z_0,z_1=z_0+f_c(z_0)=z_0+z_0^2+c, z_2=z_0+f(z_1)=z_0+(z_0+z_0^2+c)^2+c,...,
\end{equation}
different from the sequence generating the \ioj sets (see \eqref{unus})
\begin{equation*}\label{jj}
z_0,z_1=z_0^2+c,z_2=(z_0^2+c)^2+c,z_3=((z_0^2+c)^2+c)^2+c,...
\end{equation*}
For example, compare the \foj set and the \ioj obtained for $c=-0.5$ in Figs \ref{fig2} (e) and (f), respectively.

However, for $c=0$, probably the most important property is the following
\begin{proposition}
The \iom set is the \ioj set for $q\downarrow0$ and $c=0$
\begin{proof}
Consider, as required by the \foj set, $z_0$ variable within a complex lattice, and $q\downarrow0$. For $c=0$, the sequence \eqref{jj0} generating the $FO\mathcal{K}_0$, becomes
\[
z_0,z_0^2+z_0,(z_0^2+z_0)^2+z_0,...
\]
If one denote $\bar{c}=z_0$, one obtains the sequence defining the \iom set
\[
\bar{c},\bar{c}^2+\bar{c},(\bar{c}^2+\bar{c})^2+\bar{c},((\bar{c}^2+\bar{c})^2+\bar{c})^2+\bar{c},...
\]
i.e. the \iom obtained with the map $f_{\bar{c}}(z)=z^2+\bar{c}$.
\end{proof}
\end{proposition}
To note that, while the $FO\mathcal{K}_0$ set with $q\downarrow0$ for $c=0$ is identic with the \iom set (Fig. \ref{fig3} (a)), the $IO\mathcal{K}_0$ set for $c=0$ is, as known \cite{bibzece}, a filled disc (Fig. \ref{fig3} (b))

While generally it is considered that a FO continuous or discrete system for $q=1$ should identify with his own IO variant, the next result shows another surprising property of \fom map (see also. \cite{bib1,mac1,mac2,mac3} for differences between real FO systems and their IO counterparts)
\begin{proposition}\label{propi2}
For $q\uparrow 1$ the \fom set differs from the \iom set.
\begin{proof}
Consider the limit of $z(n)$ for $q\uparrow 1$, with $z(0)=0$ in \eqref{ecuss} rearranged as follows
\begin{equation*}
\lim_{q\uparrow 1}z(n)=z(n)|_{q=1}=\frac{1}{\Gamma(q)}\sum_{i=1}^{n}\frac{\Gamma(n-i+q)}{\Gamma(n-i+1)}\bigg|_{q=1}f_c(z(i-1))=\sum_{i=1}^nf_c(z(i-1)),
\end{equation*}
or, in simplified form
\[
z(n)=\sum_{i=1}^nf_c(z(i-1)),
\]
i.e., for $q\uparrow 1$, the \fom map generates a set different from the \iom set.
\end{proof}
\end{proposition}
The difference revealed by Proposition \ref{propi2} can be viewed in Figs. \ref{fig4}, where the \iom set and \fom set, for $q=1$, are presented.

\section{Echipotential lines and external rays}
\subsection{Basic notions on echipotential lines and external rays for \iom and \ioj sets}

The external arguments theory of the \iom set has been developed in \cite{mig1,mig2} and popularized in \cite{sci} and makes use of an analogy to electrodynamics. It was shown that the exterior of the \iom set can be viewed as an electrostatic field. Consider, as described in \cite{beau,mig1,mig2}, a capacitor made of a hollow metallic cylinder with a great diameter inside of which an axis of aluminum shaped in such a way that its cross-section is the Mandelbrot set. The ensembles of cylinder and axial bar are supposed to be infinitely long. In other words, one has an aluminum bar with cross section the Mandelbrot set, situated in the middle of a large hollow metallic cylinder. If the interior bar is set at potential 0 and the exterior cylinder at a high potential, between the two metallic pieces appears an electric field which creates surfaces in the surrounding space. If one considers an orthogonal  section through this ensemble of metallic corps and echipotential surfaces, one obtains the \iom set, surrounded by \emph{echipotential curves }(lemniscates), sections through the equipotential surfaces with constant potential. It has been proved that the equipotential lines are also lines of equal escape time in the time escape algorithm to generate the \iom or \ioj sets \cite{beau}.

A particle starting from the frontier of the \iom set will reach the great circle surrounding the \iom set by following the \emph{external rays}, perpendicular curve on the equipotential lines, being gradient lines of potential.

The equipotential curves are given by

\begin{equation}\label{echip}
\lim_{n\rightarrow \infty}\frac{1}{2^n}\log|f^n_c(z)|,
\end{equation}
while the external rays are defined as follows
\begin{equation}\label{line}
\lim_{n\rightarrow \infty}\frac{1}{2^n}\arg(f^n_c(z)),
\end{equation}
where, for the Mandelbrot map, $f_c(z)=z^2+c$, one has $f^0(c)=c$, $f^1_c(c)=c^2+c$,...,$f_c^n(c)=(f_c^{n-1})^2+c$,...,$n=1,2,...$

Using the Green’s function, the Douady-Hubbard potential $U(c)$ of a point $c$ situated between the cylinder and the outside of the Mandelbrot set can be written \cite{miguel1,miguel2} (see \eqref{echip})
\[
U(c)=\log|c|+\sum_{n=1}^\infty\frac{1}{2^n}\log\bigg|1+\frac{c}{{[f_c^{n-1}(c)]}^2}\bigg|.
\]
$U(c)$ is zero at points $c$ belonging to the boundary of the \iom set, while for large $c$ $U(c)$ is approximated by $\log(c)$.

An equipotential line defined by a constant $\bar{c}\in[2,\infty)$, is a closed curve surrounding the \iom set and is defined as the set of points $c$ with property: $\{c~|U(c)=\bar{c}\}$, i.e. the set
\begin{equation*}\label{bibi1}
\bigg\{c~\big|\log|c|+\sum_{n=1}^\infty\frac{1}{2^n}\log\bigg|1+\frac{c}{{[f_c^{n-1}(c)]}^2}\bigg|=\bar{c}\bigg\}
\end{equation*}

The external argument $\theta(c)$ of an external ray that pass through a point $c$, with a large $|c|$, and which determines the point where it reaches the great circle, is the argument of the function $\Phi(c)$ given as follows \cite{miguel3}
\[
\Phi(c)=c\prod_{n=1}^\infty{\bigg [1+\frac{c}{[f_c^{n-1}(c)]^2}\bigg]}^{2^{-n}}.
\]
The external argument of $\Phi(c)$ is (see \eqref{line})
\[
\theta(c)=\arg(c)+\sum_{n=1}^\infty\frac{1}{2^n}\arg\bigg[1+\frac{c}{{[f_c^{n-1}(c)]}^2}\bigg],
\]
and, therefore, the external ray for a fixed angle $\bar{\theta}\in[0,2\pi)$ is the locus of points $c$ in the complex plane that have all a same external argument with property: $\{c~|\theta(c)=\bar{\theta}\}$, i.e. the set

\begin{equation}\label{bibi2}
\bigg\{c~\big|\arg(c)+\sum_{n=1}^\infty\frac{1}{2^n}\arg\bigg[1+\frac{c}{{[f_c^{n-1}(c)]}^2}\bigg]=\bar{\theta}\bigg\}.
\end{equation}
By $\arg (c)\in[0,2\pi)$ is denoted the principal value of the argument of a complex number. To every point on the frontier of the \iom set there could exist several external rays.

\subsubsection{Approximations of equipotential lines and external rays}\label{remus}
To generate computationally equipotential curves and external rays using the above relations is a quite hard task. However, there exist several other simpler methods which can be applied both to the \iom set and \ioj sets.
\begin{itemize}
\item[1.]
In \cite{sci} the potential $U(c)$ is approximated by the value $\bar{U}(c)$ defined as following: if $|z_n|>M$, while $f_c$ is iterated, where $M$ is e.g. $10000$ \cite{sci}, the potential can be approximated by $\bar{U}(c)\approx\log|z_n|/2^n$, otherwise, i.e. $|z_n|$ remains smaller than $M$, the potential is set $0$.
If, for a considered constant $\bar{c}$, $\bar{U}(c)$ is close or equal to $\bar{c}$, then $c$, or $z_0$ in the case of Julia sets, belongs to the equipotential line and the point $c$, or $z_0$, respectively, is plotted.

An even more simple method is the Level Set Method (LSM \cite{sci}) which, to a point $c$, or $z_0$, respectively, within a complex lattice, attributes a color (e.g. black), depending to the number of iterations of $f_c$ for which $|z_n|$ remains bounded. Therefore, for each $n\in[1,N]$ one obtains a level set which, approximatively is identic to an equipotential line.

\item[2.]On the other side, the external rays \eqref{bibi2} can be approximated by the Binary Decomposition Method (BDM \cite{sci}) with respect to fixed angle $\alpha_0\in[0,2\pi)$. Thus to a point $c$, or $z_0$, respectively, within a complex lattice one attributes a color (e.g. black or white) if the argument of $z_n$, $\arg(z_n)$, belongs or not to the intervals $2^n\alpha_0\leq \arg(z_n)\leq 2^n\alpha_0+\pi(\mod 2\pi)$
\end{itemize}
Drawing precisely of an external ray, e.g. inside a detail of \iom or \fom set is strongly restricted by the number of bits of the floating-point arithmetic used by the computer program \cite{miguel1}.
Therefore, the external rays cannot be drawn in certain details with computer programs (see e.g. \cite{hh}) using the double 64 bits format.
Also, deep studies of external arguments require external arguments measured not in radians, but as fractions of complete turns. Using this unit, most of the notable points of the Mandelbrot set boundary have rational external arguments \cite{miguel1,miguel2}.

\subsection{Echipotential lines and external rays of \fom and \foj sets}

The external rays for rational angles land to the frontier of the \iom set and connected \ioj sets and, as verified numerically in this paper, this property holds for FO sets too.

Because writing analogue expressions of equipotential lines \eqref{echip} and external rays \eqref{line} for the \fom or \foj maps, called hereafter Fractional Equipotential Lines (FEL)s and Fractional External Rays (FER)s, are a difficult task, a possible overcome is to adapt the approximations given in Subsection \ref{remus} where $z_n$ is used instead $f^n_c$.

In Fig. \ref{fig5} (a) are presented the FELs of the \iom set and \fom set for $q=10^{-6}$ (see Property \ref{propi1}), and in Figs \ref{fig5} (b) and (c) are presented the FELs for \fom sets for $q=1$ and $q=0.5$ respectively.

As known, the following property of equipotential hods

\begin{proposition}\label{popo}
Equipotential lines can not intersect each other.
\end{proposition}

While for the \iom set and the \fom set, for $q=1$, this property is obviously verified computationally (Fig. \ref{fig5} (a)), for the \fom set with $q<1$ this property seems to be no more verified and the FELs intersect (see Figs \ref{fig5} (b) and (c)).

Similarly, the FELs of \foj sets seem to intersect (see Figs \ref{fig5} (d),(e),(f) where the \foj sets are determined for $q=0.5$ and correspond to $c$, chosen at the points denoted $A$, $B$ and $C$ in the \fom set in Fig. \ref{fig5} (c))

To draw FERs one divided the complex plane $\mathbb{C}$ into sectors where we set the same color if $\theta$ is within some certain interval \cite{teza2}. Another simple way to identify the external rays is to use the ordinary escape-iterations algorithm with large escape radius and plot those points $c$ for which $\Im(z_n)>0$. In Figs \ref{fig6} (a), are presented as overplot the FERs for the \iom set and \fom set for $q=10^{-6}$, in Fig \ref{fig6} (b) and (c) the FERs for \fom sets with $q=1$ and $q=0.5$ respectively and Figs \ref{fig6} (d), (e) and {f} are drawn the FERs for \foj sets corresponding to the points $A$, $B$ and $C$ taken in the \fom set in Fig. \ref{fig5} (c).

\textbf{Conflict of Interest: }The authors declare that they have no conflict of interest

\clearpage

\begin{figure}
\end{figure}

\newpage{\pagestyle{empty}\cleardoublepage}

\begin{figure}
  \includegraphics[width=.85\textwidth]{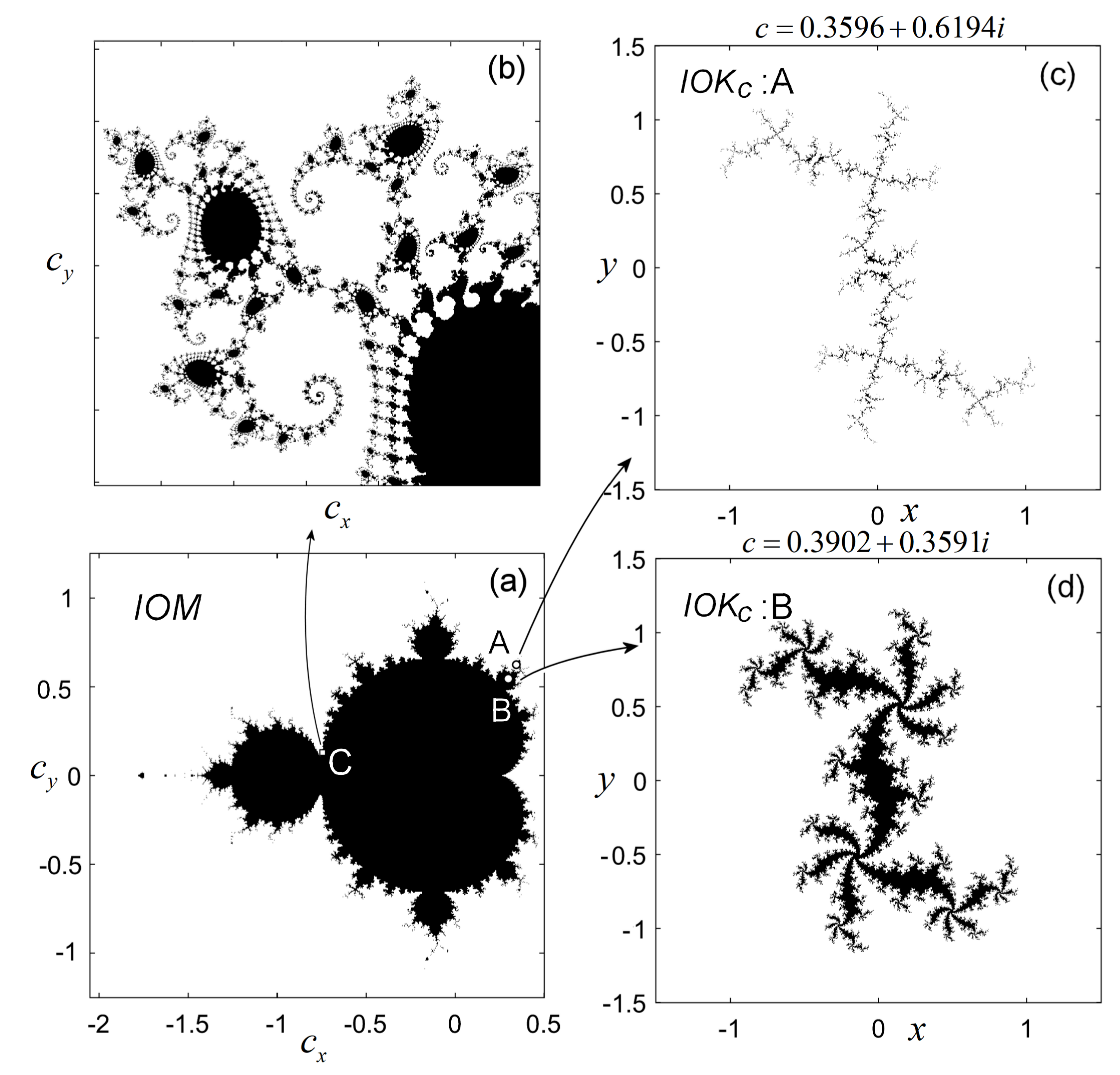}
\caption{a) \iom set; b) Zoomed detail of zone $C$ of the \iom set; c) \ioj set corresponding to $c=0.3596+0.6294i$ (point $A$ exterior to the \iom set); d) \ioj set corresponding to $c=0.3902+0.3591i$ (point $B$ from the interior of the \iom set).}
\label{fig1}       
\end{figure}

\begin{figure}
  \includegraphics[width=1\textwidth]{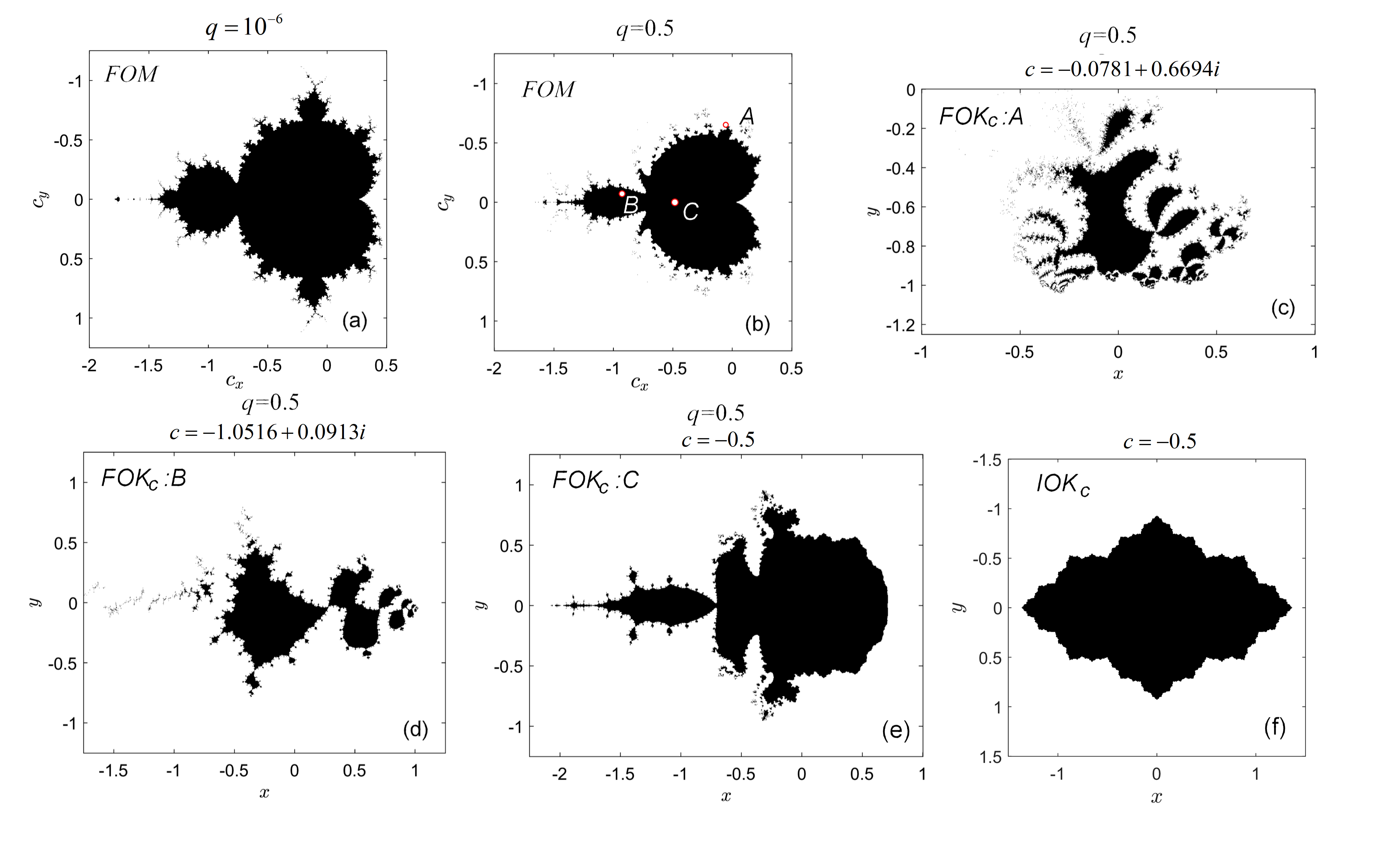}
\caption{a) \fom set for $q=10^{-6}$; b) \fom set for $q=0.5$; c) \foj set for $q=0.5$ and $c=-0.0781+0.6694i$ (see point $A$ in Fig. \ref{fig2} (b)); d) \foj set for $q=0.5$ corresponding to $c=-1.0516+0.0913i$ (see point $B$ in Fig. \ref{fig2} (b)); e) \foj set for $q=0.5$ corresponding to $c=-0.5$ (see point $C$ in Fig. \ref{fig2} (b)); f) \ioj set for $c=-0.5$.}
\label{fig2}       
\end{figure}

\begin{figure}
  \includegraphics[width=1\textwidth]{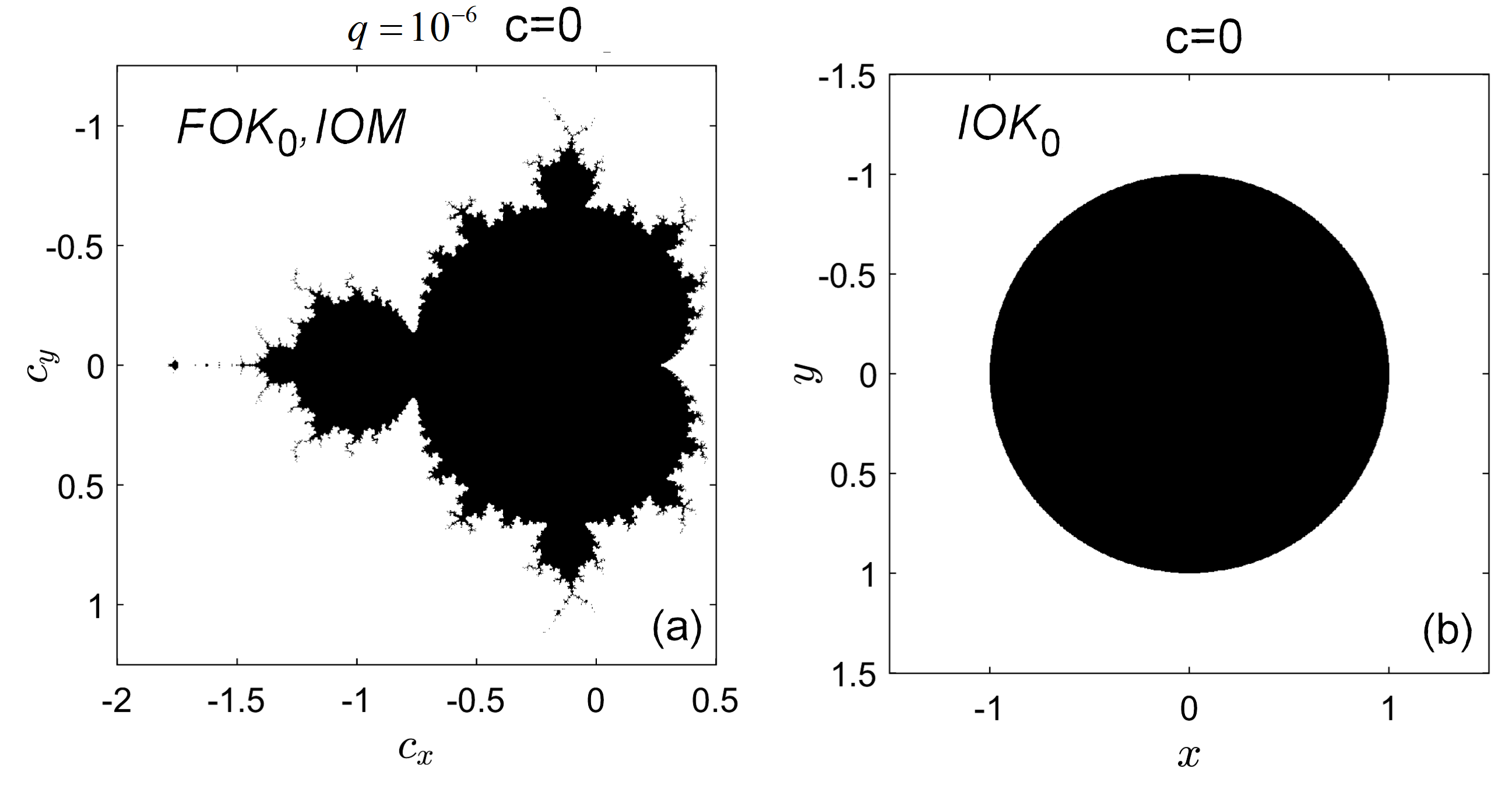}
\caption{a) The \foj set for $q\downarrow0$ and $c=0$ is identic to the \iom set; b) The \ioj set for $c=0$.}
\label{fig3}       
\end{figure}

\begin{figure}
  \includegraphics[width=1\textwidth]{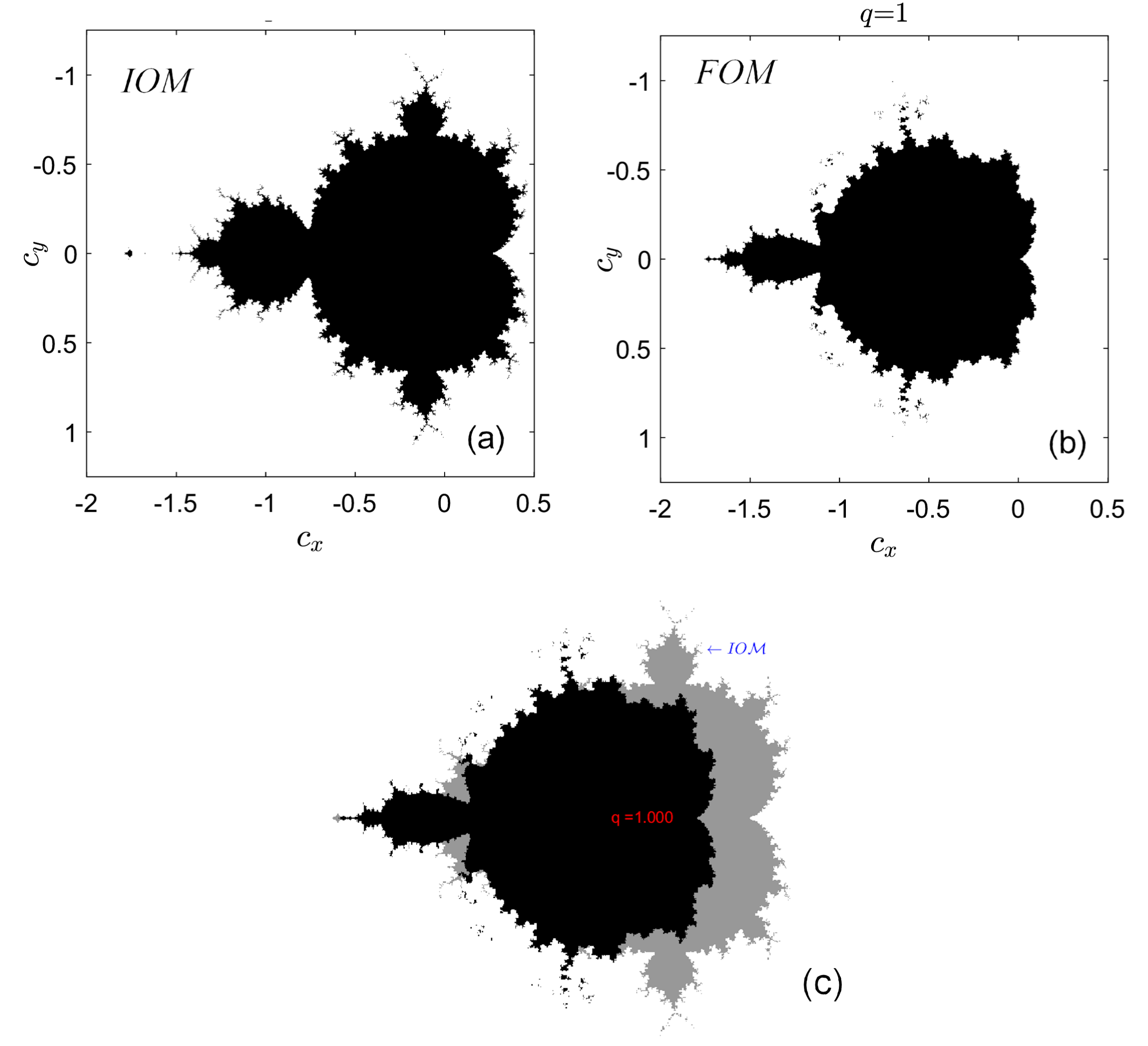}
\caption{Differences between the \iom set (a) and the \fom set for $q=1$ (b); c) Overplot of the \iom and \fom set for $q=1$ (image from animated video).}
\label{fig4}       
\end{figure}

\begin{figure}
  \includegraphics[width=1\textwidth]{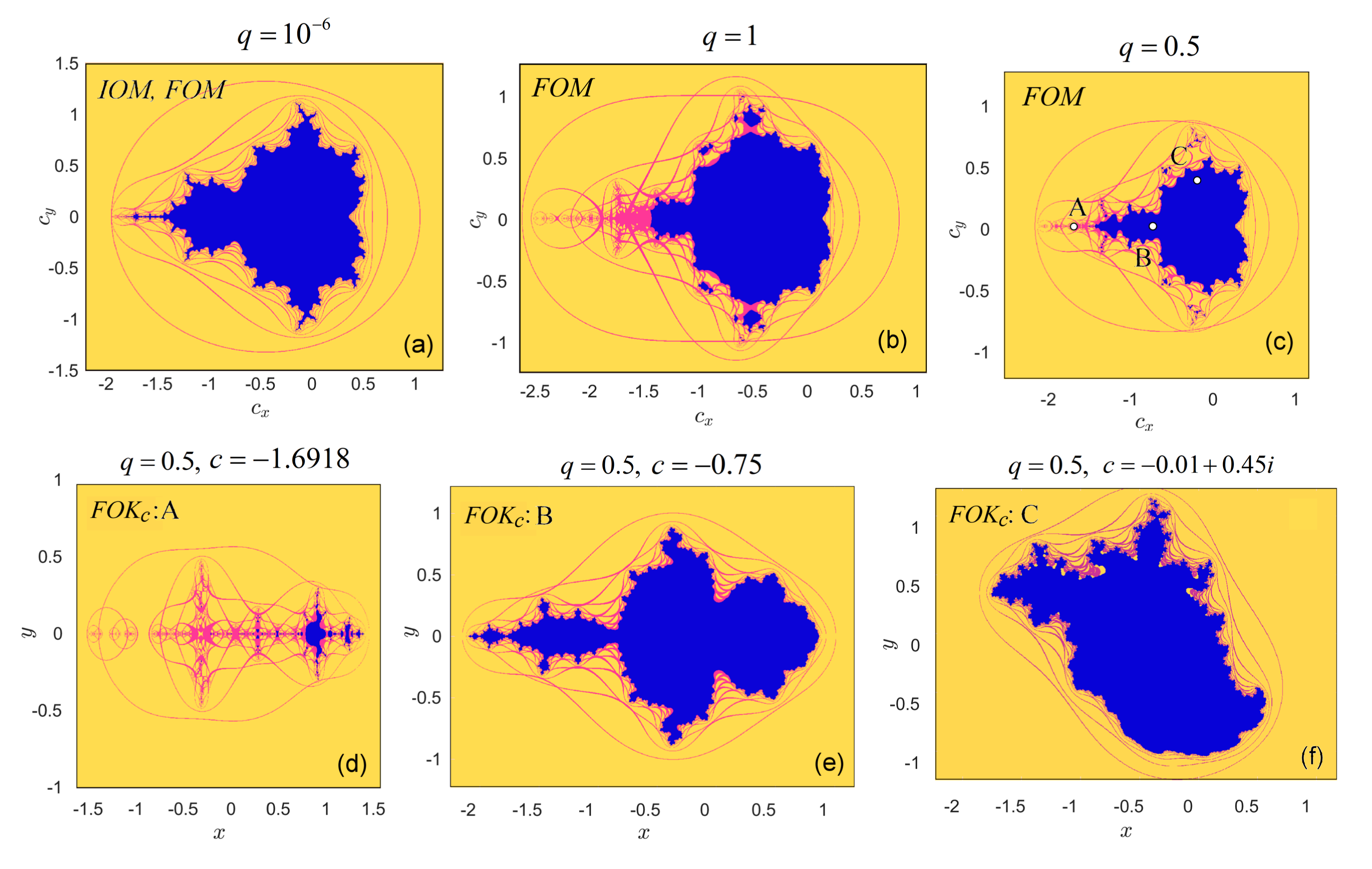}
\caption{a), b), c) FELs of three \fom sets for $q=10^{-6}$, $q=1$ and $q=0.5$, respectively; d), e), f) FELs of \foj for $q=0.5$, corresponding to points $A$, $B$, $C$ from the \fom, respectively.}
\label{fig5}       
\end{figure}

\begin{figure}
  \includegraphics[width=1\textwidth]{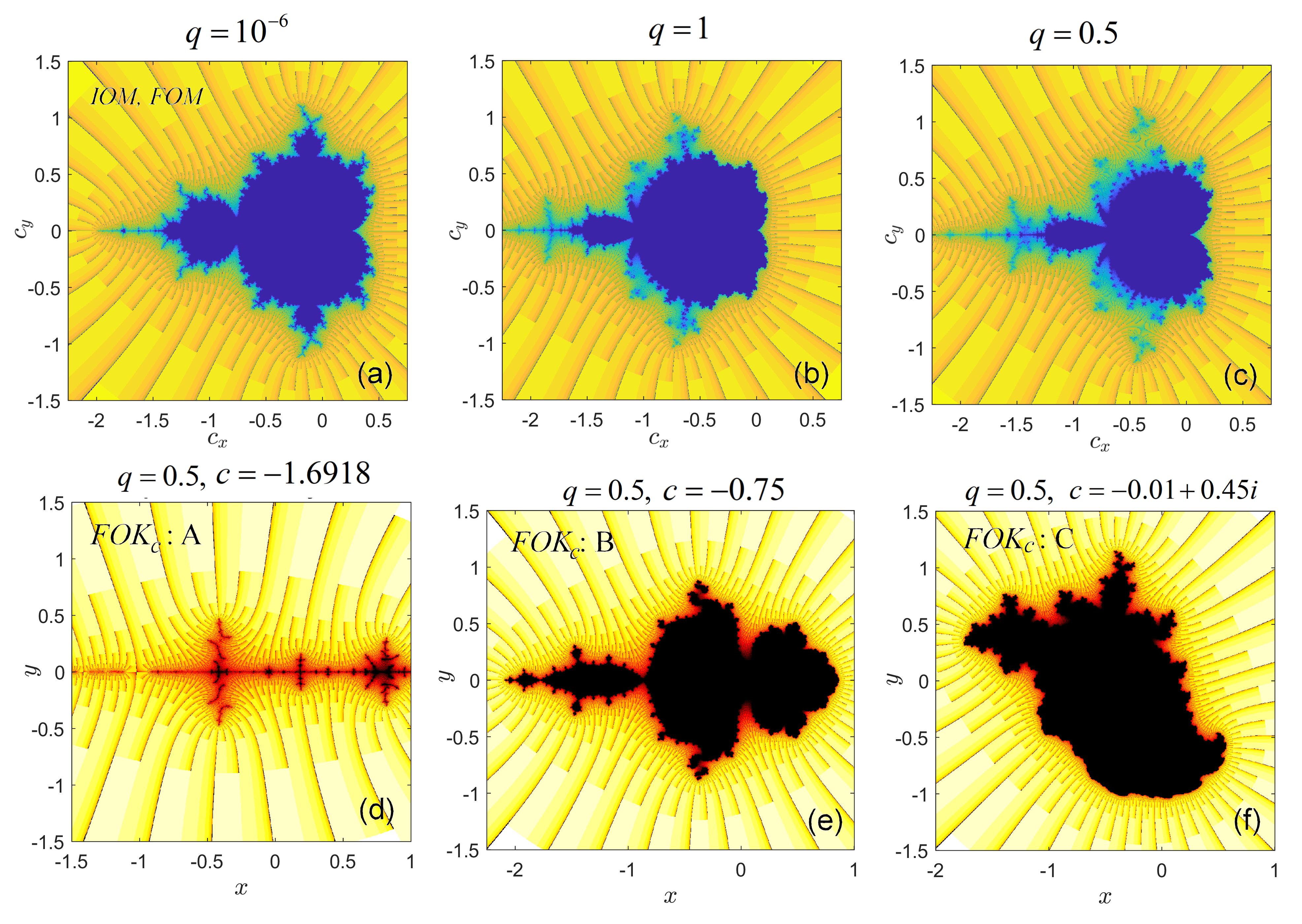}
\caption{a), b), c) FERs of three \fom sets for $q=10^{-6}$, $q=1$ and $q=0.5$, respectively; d), e), f) FERs of three \foj sets corresponding to points $a$, $B$ and $C$ (Fig. \ref{fig5} (c)),}
\label{fig6}       
\end{figure}


\clearpage

\begin{thebibliography}{100}
\bibitem{juli}Julia, Gaston (1918). M\'{e}moire sur l'it\'{e}ration des fonctions rationnelles. Journal de Math\'{e}matiques Pures et Appliqu\'{e}es (in French). 1: 47–245.
\bibitem{danca1} Danca, M.-F.; Fe\v{c}kan, M. Mandelbrot set and Julia sets of fractional order, M. Nonlinear Dynamics, 2023, 111(10), 9555–9570
\bibitem{fat}Patou, P. Sur les substitutions rationnelles. Comptes Rendus de l’Acad´emie des Sciences de Paris, 164 (1917), 806–808 and 165 (1917), 992–995,
\bibitem{oxi}Elsadany, A.A.; Aldurayhim, A.; Agiza, H.N.; Elsonbaty, A. On the Fractional-Order Complex Cosine Map: Fractal Analysis, Julia Set Control and Synchronization, Mathematics 2023, 11, 727.
\bibitem{oxi2}Fe\v{c}kan, M.; Danca, M.-F. Non-Periodicity of Complex Caputo Like Fractional Differences. Fractal Fract. 2023, 7, 68. https://doi.org/10.3390/fractalfract7010068
\bibitem{oxi3}Danca, M.-F. On the Stability Domain of a Class of Linear Systems of Fractional Order. Fractal Fract. 2023, 7, 49. https://doi.org/10.3390/fractalfract7010049
\bibitem{dancu}Danca, M.-F.; Fe\v{c}kan, M. Mandelbrot set and Julia sets of fractional order. Nonlinear Dyn 111, 9555–9570 (2023). https://doi.org/10.1007/s11071-023-08311-2

\bibitem{m1} Brooks, R.; Matelski, P. The dynamics of 2-Generator Subgroups of PSL(2,C), in Irwin Kra ed.. Riemann Surfaces and Related Topics: Proceedings of the 1978 Stony Brook Conference. 1 May (1981)


\bibitem{mana1} Mandelbrot, B.B. Fractal aspects of the iteration of z-lz(1z) for complex l, z.In: Helleman RHG, editor. Nonlinear dynamics. Annals of New York Academy of Sciences 1980;357:249–59.


	\bibitem{mand2} Mandelbrot, B.: Fractal Aspects of the Iteration of $z\mapsto z(1-z)$ for Complex $\lambda, z$. Annals of the New York Academy of Sciences. \textbf{357}(1), 249--259 (1980)

	\bibitem{bibzece} Peitgen, H.-O.; Peter, H.R.: The Beauty of Fractals Images of Complex Dynamical Systems, Springer (1986)

	\bibitem{bibunspe} Branner, B. The Mandelbrot Set. In Chaos and Fractals: The mathematics behind the computer graphics. Proc Sympos Appl Math, Vol. 39 (Ed. R. L. Devaney and L. Keen). Providence, RI: Amer. Math. Soc., 75--105, (1989)

	\bibitem{sci} Barnsley M.F.; Devaney, R.L.; Mandelbrot, B.B.; Peitgen, H.O; Saupe, D.; Voss, R.F. With Contributions by Yuval Fisher Michael McGuire. The Science of Fractal Image, Springer-Verlag, New York Inc (1988)

\bibitem{mandelus} Mandelbrot, B. The Fractal Geometry of Nature, W. H. Freeman, New York (1983)

	\bibitem{dou} Douady, A.; Hubbard, J.H. Etude Dynamique des Polynômes Complexes. Prépublications mathémathiques d'Orsay 2/4 (1984/1985)

	\bibitem{bibcinspe} Kahn, J. The Mandelbrot Set is Connected: A Topological Proof. August 2001 \url{http://www.math.brown.edu/~kahn/mconn.pdf}



	\bibitem{devu} Devaney, R. The Mandelbrot Set and the Farey Tree, and the Fibonacci Sequence. Amer. Math. Monthly \textbf{106}, 289--302 (1999)


	\bibitem{4} Diaz, J.B.; Olser T.J. Differences of Fractional Order. Math Comput \textbf{28}(125), 185--202 (1974)

	\bibitem{6} Abdeljawad T. On Riemann and Caputo Fractional Differences. Comput Math Appl \textbf{62}(3),1602--1611 (2011)

	\bibitem{mich} Fe\v{c}kan, M; Posp\'i\v sil, M.; Danca, M.-F.; Wang, J. Caputo Delta Weakly Fractional Difference Equations. Fract Calc Appl Anal 25, 2222–-2240 (2022)



	\bibitem{9} Cermak, J.; Gyori, I.; Nechvatal, L. On Explicit Stability Conditions for a Linear Fractional Difference System. Fract Calc Appl Anal \textbf{18}, 651--672 (2015)

	\bibitem{10} Chen, F.L. A review of Existence and Stability Results for Discrete Fractional Equations. J Comput Complex Appl \textbf{1},22--53 (2015)

	\bibitem{x1} Danca, M.-F. Symmetry-Breaking and Bifurcation Diagrams of Fractional-order Maps. Commun Nonlinear Sci \textbf{116}, 106760 (2023)

	\bibitem{x2} Danca, M.-F. Fractional Order Logistic Map: Numerical Approach. Chaos Soliton Fract \textbf{157}, 111851 (2022)

	\bibitem{bib1} Danca, M.-F.; Fe\v{c}kan, M., Kuznetsov, N.; Chen, G.: Coupled Discrete Fractional-Order Logistic Maps. Mathematics \textbf{9}, 220 (2021)

	\bibitem{micy} Dibl\' ik, J.; Fe\v ckan, M.; Posp\'i\v sil, M. Nonexistence of Periodic Solutions and $S$-Asymptotically Periodic Solutions in Fractional Difference Equations. Appl Math Comp \textbf{257}, 230--240 (2015)

	\bibitem{wiki} \url{https://en.wikipedia.org/wiki/Mandelbrot_set#cite_note-bf-4}

\bibitem{cod}\url{https://www.mathworks.com/matlabcentral/fileexchange/121632-fo_mandelbrot}

	\bibitem{mil} Milnor, J. Dynamics in One Complex Variable. Third Edition. (AM-160). Princeton University Press, (2006)





	\bibitem{bibnoua} Atici, F.M.; Eloe, P.W. Initial Value Problems in Discrete Fractional Calculus. Proc Amer Math \textbf{137}(3), 981-–989 (2009)

	\bibitem{bibcinci} Anastassiou, G. Principles of Delta Fractional Calculus on Time Scales and Inequalities. Math Comput Model \textbf{52}(3--4), 556--566 (2010)

	\bibitem{time} Agarwal, R.P.; Bohner, M.: Basic Calculus on Time Scales and Some of its Applications. Results Math \textbf{35}(1--2), 3--22 (1999)

\bibitem{plu1}Magin, R. L.; Ovadia, M.: Modeling the cardiac tissue electrode interface using fractional calculus. J Vibr Control 14(9--10), 1431–-1442 (2008)
\bibitem{plu2}Heymans, N.: Dynamic measurements in long-memory materials: fractional calculus evaluation of approach to steady state. J Vibr Control \textbf{14}(9--10), 1587-–1596 (2008)
\bibitem{plu4}Lima, M.F.M.; Machado, J. A. T.; Cris\'{o}stomo, M.: Experimental signal analysis of robot impacts in a fractional calculus perspective. Journal of Advanced Computational Intelligence and Intelligent Informatics, \textbf{11}9, 1079–-1085 (2007)
\bibitem{plu5}Debnath, L.: Recent applications of fractional calculus to science and engineering. International Journal of Mathematics and Mathematical Sciences, \textbf{2003}(54), 3413–-3442 (2003)
\bibitem{plu15} Wang, Y.; Li X.; Wang, D.; Liu, S.: A brief note on fractal dynamics of fractional Mandelbrot sets. Appl Math Comput 432, 127353 (2022)
\bibitem{mumu}Huang, L.; Park, J.H.; Wu, G.C.; Mo, Z.W.: Variable-order fractional discrete-time recurrent neural networks. J Comput App Math 370 112633 (2020)
\bibitem{mumu2}Wu, G.C. Baleanu; D. Lin, Z.X.: Image encryption technique based on fractional chaotic time series, J Vibr Contr 22, 2092--2099 (2016).
\bibitem{mumu3}Abdeljawad, T.; Banerjee, S.; Wu, G.C.: Discrete tempered fractional calculus for new chaotic systems with short memory and image encryption. Optik 203, 163698 (2020)
\bibitem{ur}\url{https://www.math.univ-toulouse.fr/~cheritat/wiki-draw/index.php/Mandelbrot_set}
\bibitem{mac1}Danca, M.-F.; Kuznetsov, N. D3 Dihedral Logistic Map of Fractional Order, Mathematics 2022, 10, 213. https://doi.org/10.3390/math10020213
\bibitem{mac2}Danca, M.-F. Fractional order logistic map: Numerical approach, Chaos, Solitons \& Fractals, 157, 2022, 111851
\bibitem{mac3}Danca, M.-F. Symmetry-breaking and bifurcation diagrams of fractional-order maps, Communications in Nonlinear Science and Numerical Simulation, 116, 106760, 2023
\bibitem{mig1}Douady, A.; Hubbard, J.H. It\'{e}ration des polyn\^{o}mes quadratiques complexes,Comptes Rendus des S\'{e}ances de l’Acad\'{e}mie des Sciences. S\'{e}rie I. Math\'{e}matique, vol. 294, no. 3, pp. 123–126,
1982.
\bibitem{mig2}Douady, A. Algorithms for computing angles in the Mandelbrot set, in Chaotic Dynamics and Fractals, M. Barnsley and S. G.Demko, Eds., vol. 2, pp. 155–168, Academic Press, NewYork, NY, USA, 1986.
\bibitem{teza2} Thomas G. de Jong, Dynamics of Chaotic Systems and Fractals, Bachelor Thesis in Applied Mathematics, University of Groningen, 2009
\bibitem{miguel1}Pastor, G.; Romera, M., \'{A}lvarez, G.; Montoya, F. Operating with external arguments in the Mandelbrot set antenna, Physica D 171 (2002) 52–71.
\bibitem{miguel2}Romera, M.; Pastor, G.; Orue, A.B.; Martin, A.; Danca, M.-F.; Montoya, F. A Method to Solve the Limitations in Drawing External Rays of the Mandelbrot Set, Mathematical Problems in Engineering
Volume 2013, Article ID 105283, 9 pages http://dx.doi.org/10.1155/2013/105283
\bibitem{miguel3}Carleson L.; Gamelin, T.W. Complex Dynamics (Springer, New York, 1993), 139.
\bibitem{hh}Jung, W. “Mandel: software for real and complex dynamics,” 2012, http://www.mndynamics.com/indexp.html.
\bibitem{beau}Peitgen, H.O.; Richter, F.H.The Beauty of Fractals, Images of Complex Dynamical Systems, Springer-Verlag, 1986


\end{thebibliography}
\end{document}